# Anomalous magnetoresistance and magnetocaloric properties of NdRu$_2$Ge$_2$

Bibekananda Maji[1], K. G. Suresh[1,*] and A. K. Nigam[2]

[1]Department of Physics,

Indian Institute of Technology Bombay, Mumbai- 400076, India

[2]Tata Institute of Fundamental Research,

Homi Bhabha Road, Mumbai- 400005, India

Abstract

It is found that the polycrystalline NdRu$_2$Ge$_2$ undergoes two successive magnetic transitions at $T_t$=10 K and $T_N$=19 K. Evidence of metamagnetic transition is detected in the magnetization isotherm data in the antiferromagnetic regime. Temperature dependence of magnetoresistance (MR) show that the relative magnitudes of MR at $T_N$ and $T_t$ change considerably as the field is increased from 10 kOe to 30 kOe. Moreover, the MR is found to be positive below 9 K for 30 kOe field although the material is ferromagnetic at these temperatures. The highest value of negative MR near $T_N$ is about 42% in a field of 30 kOe, while the positive MR is about 35 % at 3 K in a field of 50 kOe. Like MR, the magnetocaloric effect at $T_N$ and $T_t$ also shows anomalous behavior. The relative magnitudes of MCE at these temperatures are found to change with increase in field. It appears that the high field (>10 kOe) magnetic state below $T_N$ is complex, giving rise to some antiferromagnetic-like fluctuations, affecting the MR and MCE behavior.

Key words: intermetallics, magnetoresistance, magnetocaloric effect

PACS: 75.47.De, 75.30.Sg, 75.47.Np

*Corresponding author (email: suresh@phy.iitb.ac.in)

**I. Introduction**

Rare earth (R) – transition metal (T) intermetallics represent one of the most important classes of magnetic materials that give rise to many interesting properties[1-3]. Magneto-transport and magneto-thermal properties of these materials are of special interest because of the insight that they give towards the underlying magnetism of these materials. These two phenomena are also important because of the possible exploitation in applications. Therefore, the search for novel and potential materials which exhibit enhanced magneto-transport and magneto-thermal properties is on for a long time. It has been found that there is a strong correlation between these two properties in many materials. The most obvious manifestations of these two properties are the magnetoresistance (MR) and the magnetocaloric effect (MCE), respectively.

The discovery of giant magnetoresistance (GMR) in magnetic multilayer system Fe/Cr/Fe[5,6] has attracted the attention of a lot of researchers all over the world towards the study of magnetoelectronics. The spin dependent scattering at the interface between ferromagnetic and nonmagnetic layers is believed to be responsible for GMR in multilayer systems. GMR is not only seen in magnetic multilayers but in single layer thin films and bulk materials also.[6-10] Existence of large MR is also observed in a large number of bulk intermetallic compounds.[11-17] Many of these intermetallic compounds are naturally occurring layered compounds and the behavior of MR is highly anomalous in them.[12-14,16,17] Interestingly, many such materials also show considerable MCE.

The ternary intermetallic compounds $RM_2X_2$ (M=Mn/Ru, X=Si/Ge)[18-21] have long been identified as a naturally occurring layered system and offers the possibility to study the interplay between the structure and magnetism. This is because some interatomic distances (M-M) are close to the critical bond lengths corresponding to the boundary between ferromagnetic (FM) and antiferromagnetic (AFM) interactions. These compounds crystallize in the body-centered tetragonal $ThCr_2Si_2$ type structure (space group I4/mmm), where R, M, and X atoms occupy 2(a), 4(d), and 4(e) sites respectively and form repeated layers along the c-axis, stacked in the sequence R-X-M-X-R. The crystal structure and the magnetic properties of $RRu_2Ge_2$ compounds, which belong to this family, was first studied by Felner and Nowik.[20] According to their study, $RRu_2Ge_2$ compounds containing light rare earths (R= Ce to Eu) show the characteristics of ferromagnets, whereas those with heavy rare earths (R= Gd, Tb, Dy, Ho) order antiferromagnetically and undergo a metamagnetic transition in relatively low magnetic fields (< 10 kOe). In contrast to the other members of the series, $NdRu_2Ge_2$ exhibits two magnetic phase transitions, at 17 K and at 10 K.[20,22,23] The neutron diffraction study on this compound has confirmed that it enters the AFM state below 17 K ($T_N$) and this AFM state is replaced by FM state through a first order AFM to FM transition at 10 K ($T_t$).[22] Since both MR as well as MCE strongly depend on the magnetic state, we have decided to carry out a detailed magnetoresistance and magnetocaloric study, the results of which are presented in this paper.

## II. Experimental details

The polycrystalline sample of $NdRu_2Ge_2$ was prepared by arc melting a stoichiometric mixture of Nd (99.9 - at. % purity), Ru (99.9- at. % purity) and Ge (99.999-at. % purity) in a water-cooled copper hearth under high purity argon atmosphere. The resulting ingot was turned upside down and remelted four times to ensure homogeneity. The weight loss after the final melting was ~ 0.4 %. The arc melted ingot was sealed in an evacuated quartz tube and annealed at 800 °C for a week. The structural analysis of the sample was performed by the room temperature powder x-ray diffractogram (XRD) taken using Cu-K$_\alpha$ radiation. The magnetization measurements were carried out using a SQUID VSM (Quantum Design). The electrical resistance measurements were performed by a conventional four-probe method in the physical property measurements system (Quantum Design, PPMS-6500).

## III. Results and discussion

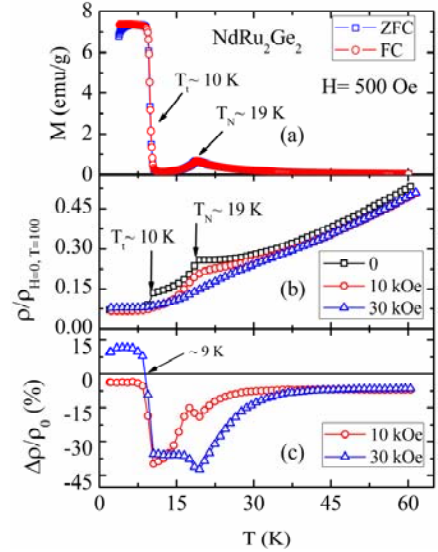

FIG. 1. (a) Temperature variation of ZFC and FC magnetization of $NdRu_2Ge_2$ in 500 Oe. (b) Normalized electrical resistivity measured as a function of temperature in the presence of different magnetic fields. (c) Temperature dependence of magnetoresistance ($\Delta\rho/\rho_0$) at 10 kOe and 30 kOe.

Fig. 1 (a) shows the temperature variation of magnetization of $NdRu_2Ge_2$ at 500 Oe in the zero field cooled (ZFC) and field cooled (FC) modes. The transitions at $T_N$=17 K and at $T_t$=10 K, mentioned earlier, can be seen in this figure. However, as the field is increased to 10 kOe, the transition at $T_t$ disappears completely and the magnetization at the lowest temperature becomes almost independent for further increase in field. Temperature dependence of electrical resistivity normalized to the value at 100 K in the absence of a field as well as in the presence of different fields (10 kOe and 30 kOe) is shown in Fig.1 (b). Two distinct kinks at close to 19 K and at 10 K reflect the transitions seen in the M-T data. The transition to the ferromagnetic phase (~10 K) is not very much prominent in the data taken in the presence of a field. In the paramagnetic region, the resistivity varies almost linearly with temperature, indicating that the electron-phonon interaction is dominant.

The temperature dependence of magnetoresistance defined as $MR(\%) = \left(\frac{\rho(H) - \rho(H=0)}{\rho(H=0)}\right) \times 100$, in fields of 10 kOe and 30 kOe is plotted in Fig. 1 (c). Though the MR is negative throughout the temperature range investigated for H=10 kOe, there is a sharp decrease in the magnitude at about 9 K. As the field is increased to 30 kOe, the MR exhibits a sign reversal at about 9 K. The negative sign of the MR changes to positive at this temperature and remains positive down to the lowest temperature of measurement (i.e., 3 K). The highest value of the positive MR is found to be about 12 %. It is surprising to see such a large positive MR in this temperature regime, though the compound is in the ferromagnetic state, after undergoing a AFM-FM transition (on cooling) at 10 K.

One can see that well into the paramagnetic region, the MR is almost identical for both the fields and is negative as expected. The MR data clearly reflect the M-T data by exhibiting minima at 19 K and 10 K corresponding to the antiferromagnetic and ferromagnetic transitions respectively. The MR corresponding to these minima is found to be 19 % and 40 % at 19 K and 10 K respectively for H=10 kOe. These values respectively change to 42 % and 35 % at 30 kOe. It is important to notice that when the field is increased from 10 kOe to 30 kOe, it has opposite effects on the two transitions. This dramatic increase in MR (or a decrease in the resistivity) at $T_N$ as the field is increased from 10 to 30 kOe suggests an increase in the ferromagnetic ordering with increase in field. However, one cannot rule out the role of any magneto-structural contribution for this change. More importantly, the magnitude of (negative) MR near the ferromagnetic transition has decreased slightly as the field is increased, which indicates the onset of some positive contribution to MR at this temperature. Interestingly, the MR at the lowest temperature is found to be nearly zero at 10 kOe and strongly positive at 30 kOe. MR isotherms measured in the range of -50 to +50 kOe at different temperatures are shown in Fig. 2(a). Before recording the data, each temperature was achieved by cooling the sample in zero field from the paramagnetic state. One can see that both at 3 K and at 8 K, the MR is positive in the entire range except for a narrow low field region, in agreement with the MR vs. T data. It can be seen that at these temperatures, the low field (up to 5 and 7 kOe for 3 K and 8 K data, respectively) MR is nominally negative. Moreover, the positive MR continues to increase linearly with field and reaches about 36% and 31% at 3 K and 8 K respectively at 50 kOe. As the temperature is raised to 15 K, the nature of MR isotherms changes abruptly. It starts from 0 and increases to +5 % at 4 kOe, then decreases and changes the sign to reach a value of about -30% in the high field region. The behavior is found to be identical for the positive and the negative field variations. It is well known that the resistivity of an antiferromagnet decreases if the antiferromagnetic configuration is forced to a ferromagnetic configuration by the application of the field. Therefore when the field is applied to the system in its antiferromagnetic region, it can switch into the ferromagnetic state (via a metamagnetic transition) at a critical magnetic field. The field dependence of magnetization at 15 K for NdRu$_2$Ge$_2$ shows metamagnetic transition at a critical field (H$_c$) of ~4.4 kOe (see Fig.2 b). Therefore the negative contribution to MR seen at this temperature for the high field region is consistent with the magnetization data. The initial increase of positive MR as the field approaches the critical field can be attributed to the enhanced scattering due to intersite fluctuations.[11] Once the critical field is overcome the ferromagnetic component grows rapidly with further increase in field, which causes MR to increase sharply to -31 % at about 20 kOe. The magnitude of MR then decreases almost linearly with a negligible slope in the higher field (>30 kOe) regime. This slow decrease of MR at higher fields indicates the presence of an additional positive contribution. This positive linear behavior is found to be the common feature of antiferromagnetic R$_2$Ni$_3$Si$_5$ compounds.[12,24] Antiferromagnetic PrCu$_2$Si$_2$ was also found to exhibit the same feature.[25] Since, the structure of R$_2$Ni$_3$Si$_5$ has RNi$_2$Si$_2$ as the building block, it was assumed that this linear term may arise from the RM$_2$X$_2$ structure.

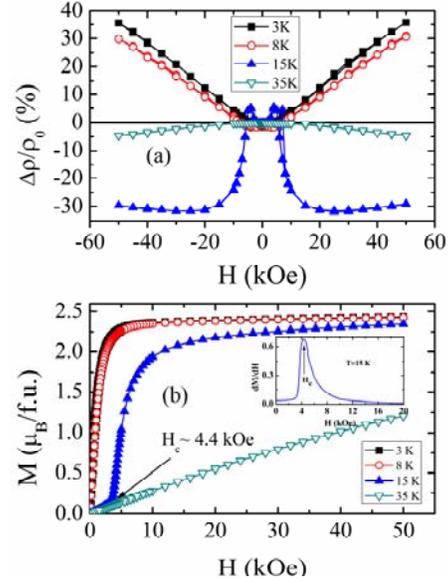

FIG. 2. (a) Magnetoresistance as a function of magnetic field (H) measured at some selected temperatures for $NdRu_2Ge_2$. (b) Magnetization (M) as a function of magnetic field (H) at some selected temperatures. Inset shows the derivative of the *dM/dH* vs. *H* at 15 K, indicating the occurrence of metamagnetic transition at $H_C \sim$ 4.4 kOe .

From the data presented above, one of the key findings is that the MR is positive below 9 K despite the ferromagnetic ordering at these temperatures. In general, the MR is expected to be negative in compounds with a long range ferromagnetic order because of the suppression of spin fluctuations by the external field. However, relatively large positive MR is reported in various antiferromagnetically ordered compounds.[12-14,26] Yamada and Takada[27] have shown theoretically that MR is positive when the applied field is parallel to the sublattice magnetization. The authors have also predicted that the MR of a polycrystalline antiferromagnetic material can change from positive to negative as the magnetic field is increased. Such a behavior is seen in the present data at 15 K. The positive MR is usually understood in terms of the Lorentz force contribution. A large positive MR, even as high as a few hundred per cent, is observed in pure metals[28] and single crystals when $\omega_c \tau >> 1$, where $\omega_c$ and $\tau$ are the cyclotron frequency and relaxation time of the conduction electrons respectively. The above condition is found to be satisfied in many pure elemental metals at low temperatures (where $\tau$ is very large and therefore the resistivity is very small, $\sim n\Omega cm$). But the Lorentz force cannot be the origin of such a large positive MR in the present compound because the residual resistivity (at 3 K) is about 10 $\mu\Omega cm$, which implies $\tau$ is quite small. A positive MR was reported in ferromagnetic compounds UCu$_2$Ge$_2$,[29] CeFe$_2$,[30] and Fe$_{1-x}$Co$_x$Si,[31] where the authors have attributed this to the possible antiferromagnetic fluctuations at low temperatures. A large positive MR has also been reported in isostructural, naturally occurring multilayer compounds SmMn$_2$Ge$_2$ and LaMn$_2$Ge$_2$, in the ferromagnetic state.[14,17] The scattering of the conduction electrons by the nonmagnetic layers and the presence of a field induced pseudo-gaps in some portion of the Fermi surface was assumed to be the possible reason for the positive MR. A similar prediction can also be made for the present compound. Gerber et al.[32] have shown that isotropic positive MR with linear field dependence is an inherent property of geometrically constrained ferromagnets. They claimed that it is the quantum electron-electron interference effect which results in positive and linear MR. Considering the fact that some of the isostructural materials[14,17], which are considered to be the naturally occurring multilayered compounds exhibit a positive MR in the ferromagnetic state, we feel that layered structure of the present compound is plying a crucial role in determining the transport behavior in presence of a field. In fact, the large positive MR in R$_2$Ni$_3$Si$_5$ [13] was understood on the basis of their layered structure similar to Dy/Sc superlattices[33] in which giant positive MR seems to arise from the multiple reflections of carriers from the interfaces before scattering, which enhances the sensitivity of resistance to momentum loss due to reflection. In fact these reasons also explain the variation of MR at T$_t$ as the field is increased from 10 to 30 kOe (shown in Fig. 1c). Another reason for the large low temperature positive MR is the domain wall pinning, which is usually prominent in materials with large anisotropy. Since the magnetic anisotropy in the present case is seen to be small (as evident from the M-H pots), it appears that this contribution is insignificant.

Fig. 3 presents the Arrott plots for NdRu$_2$Ge$_2$ in the temperature range 3-29 K and its inset shows the low field region in the temperature range of 11–17 K. According to Banerjee criterion [34, 35], negative slope in the H/M versus M$^2$ curves indicates that the magnetic phase transition is of first order, while a positive slope suggests that it is of second order.

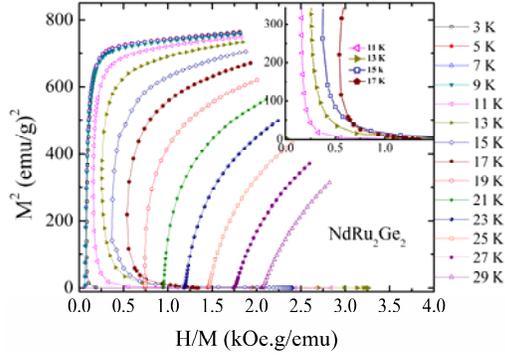

FIG. 3. Arrott plots of $NdRu_2Ge_2$ from 3 to 29 K with a temperature step of 2 K. The inset shows low field Arrott plots in the temperature range of 11–17 K.

As can be seen from the inset of Fig. 3, the slopes are clearly negative in the AFM state ($T_t < T < T_N$), which indicates the existence of field induced first order AFM-FM transition at these temperatures.

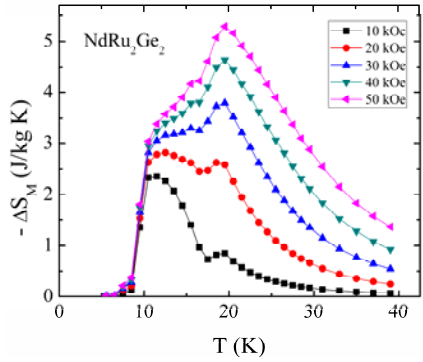

Fig. 4. Temperature variation of isothermal magnetic entropy change in $NdRu_2Ge_2$ in different fields

Since, like MR, the magnetocaloric effect measured in terms of isothermal magnetic entropy change ($-\Delta S_M$) is also critically dependent on the magnetic state of the material, we have measured the entropy change associated with the two magnetic transitions. For this we have used the Maxwell's relation, namely[36]

$$\Delta S_M(T, \Delta H) = \int_{H_1}^{H_2} \left( \frac{\partial M(T,H)}{\partial T} \right)_H dH \ldots\ldots(1)$$

Fig. 4 shows the temperature variation of the entropy change ($\Delta S_M$) in different field changes. As can be seen, for low fields (10 and 20 kOe), there are two well defined peaks, corresponding to the two magnetic transitions. Both the peak positions coincide with the transition temperatures. At 10 kOe, the peak value of the entropy change is lower at $T_N$ as compared to that at $T_t$. As the field increases to 30 kOe or more, the peak at $T_N$ becomes more prominent and its height overtakes that at $T_t$. This implies that the entropy change at $T_t$ does not increase considerably with increase in the field. Antiferromagnetic fluctuations mentioned above may be the one of the reasons for this trend. Furthermore, the MCE peak at $T_N$ is found to be quite broad, indicating that the forced ferromagnetic state between $T_N$ and $T_t$ is not strictly collinear. As reported by the neutron diffraction studies[22], the low temperature magnetic state of this compound is sine wave modulated, which is in agreement with the MCE data. This may also be the reason for the moderate MCE value in spite of the fact that the magnetic transition at $T_t$ is quite sharp (Fig.1 a).

Therefore, it is clear that the field and temperature dependencies of magnetoresistance and magnetocaloric effect $NdRu_2Ge_2$ are quite anomalous. The observation of large positive MR and the fact that the entropy change at $T_C$ does not scale with field strongly indicate the presence of a magnetic contribution behind these anomalies. It appears that the high field (>10 kOe) magnetic state below $T_N$ is complex, giving rise to some antiferromagnetic-like fluctuations, affecting the MR and MCE behavior. Additional contributions arising from the peculiar structure of the material seem to result in a large positive MR. In summary, we show that $NdRu_2Ge_2$ presents an interesting scenario as far as magneto-transport and magneto-thermal behavior are concerned.

**Acknowledgement**

The authors thank D. Buddhikot, TIFR for his help in the resistivity measureemnts.